\documentstyle[aps,prl,multicol,epsfig,amssymb]{revtex}

\begin{document}

\title{Analysis of the Metallic Phase of Two-Dimensional Holes in 
SiGe in Terms of Temperature Dependent Screening}

\author{V. Senz$^1$, T. Ihn$^1$, T. Heinzel$^1$, K. Ensslin$^1$,
G. Dehlinger$^2$, D. Gr\"utzmacher$^2$, U. Gennser$^2$}

\address{$^1$ Laboratory of Solid State Physics, ETH Z\"urich, CH-8093 
Z\"urich,
$^2$ Paul Scherrer Institut, CH-5232 Villigen PSI}

\date{\today}

\maketitle

\begin{abstract}
    We find that temperature dependent screening can quantitatively explain
    the metallic behaviour of the resistivity on the metallic side of the 
    so-called metal-insulator transition in p-SiGe. Interference and 
    interaction effects
    exhibit the usual insulating behaviour which is expected 
    to overpower the metallic background at sufficiently low 
    temperatures. We find empirically that the concept of a Fermi-liquid
    describes our data in spite of the large $r_{s}\approx 8$.
\end{abstract}

\begin{multicols}{2}
\narrowtext

%According to the quantum mechanical concept of metals and insulators
%introduced by Wilson \cite{metal}, metallic behaviour is 
%expected for systems in which at zero temperature the density of states
%at the Fermi-level remains finite. It was realized later that this 
%condition is not sufficient
Conductivity measurements give experimental access to the nature of 
the electronic ground state of disordered conductors.
At zero temperature ($T=0$) extended wave 
functions at the Fermi-level result in metallic behaviour (zero or 
finite resistivity $\rho$), while localised
wave-functions lead 
to insulating properties (infinite resistivity) \cite{anderson}.
%It can be shown theoretically that localised wave functions 
%lead to zero conductivity while extended wave functions lead to zero 
%or finite resistivity at zero temperature. 
%Measurements of the resistivity have proven to give 
%experimental access to the nature of the ground state at low 
%temperatures.
Since in experiments the 
absolute zero is not accessible, the criterion
$d\rho/dT >0$ ($d\rho/dT <0$) measured at the lowest accessible temperatures
is thought to indicate a metallic (insulating) ground state.
%while 
%should hint towards an insulating ground state.

In two-dimensional (2D) systems it was suggested theoretically 
\cite{anderson1} and supported experimentally \cite{exp,poole} that the ground 
state at zero temperature is of insulating nature. This 
belief has been challenged by the interpretation of experiments
on high mobility Si-MOSFETs \cite{kravchenko}. Later on, experiments 
on GaAs hole \cite{shayegan,cambridge,hamilton} and electron gases 
\cite{ribeiro}, AlAs electron 
gases and SiGe 
hole gases \cite{coleridge,senz,senz2} were shown to exhibit similar features as observed 
in the Si-MOSFET systems \cite{review}. As a consequence, the existence 
of a metal-insulator transition (MIT) in 2D-systems
%at zero magnetic field
has been
%intensely and
controversially discussed 
\cite{metint}.

In this paper we discuss the metallic behaviour observed in our 
p-SiGe quantum wells \cite{senz}. 
%Magnetotransport data taken across the MIT are analysed.
Interference 
corrections to the conductivity are extracted from weak localisation 
studies at low magnetic fields and interaction corrections to the 
conductivity are obtained from the temperature dependence of the 
Hall resistivity, similar to Ref. \cite{cambridge}.
In addition, following the suggestion in Ref. 
\onlinecite{dasSarma}, we compare the remaining 
(Drude) part of the conductivity with the theory for temperature 
dependent screening suitable for our system \cite{dolgopolov}.
We arrive at the following conclusions: (I) In the 
metallic regime the 2D-hole 
gas in our SiGe samples behaves like an ordinary Fermi liquid and 
exhibits localising interference and interaction corrections to the conductivity 
which can be described by conventional theory. (II) The metallic 
temperature dependence of the resistivity can be described by 
the theory of temperature dependent screening entering the 
Drude part of the resistivity.

Our samples were grown by MBE (molecular beam epitaxy).
They comprise a 20 nm 
Si$_{0.85}$Ge$_{0.15}$ quantum well where the two-dimensional hole gas 
is formed, sandwiched between undoped Si layers. The structures are 
remotely doped with boron at a distance of 15 nm above the quantum well
and gated with a Ti/Al Schottky gate.
For more details see Ref. \onlinecite{senz}.
Measurements were carried out at 
temperatures between 180 mK and 15 K using four-terminal AC 
and DC techniques.

The SiGe quantum well is compressively strained normal to the growth 
direction, which leads, together with the confinement, to a splitting 
between heavy-hole (HH) and light-hole (LH) bands of
the order of 25 meV \cite{emeleus}. 
The 2D hole gas resides in the lowest HH subband with effective 
mass $m^{\star}=0.25 m_{0}$ as determined from Shubnikov-de Haas 
oscillations.
%The in-plane LH mass is predicted to be larger than the 
%HH mass \cite{whall}.
The hole density in ungated areas of 
the device was $4.3\times 10^{11}$ cm$^{-2}$.
With the gate the hole density could be tuned between 
$1.1-2.6\times 10^{11}$cm$^{-2}$, i.e. the Fermi energy was below 2 meV.
The hole mobility in these structures
increases with carrier concentration from 1000 cm$^2$/Vs at the 
lowest to 7800 cm$^2$/Vs at the highest density similar to other p-SiGe
structures \cite{whall}. The ratio of the Drude scattering time $\tau_{e}$
and the quantum lifetime $\tau_{q}$ extracted from Shubnikov-de Haas 
oscillations is close to one indicating that transport is dominated by 
short range scattering potentials. It was confirmed in other studies 
that large angle scattering at interface charges dominates the 
mobility in such structures \cite{emeleus,whall}.

Recently Coleridge et al. \cite{coleridge} and we \cite{senz,senz2} have shown 
that p-SiGe exhibits the 
characteristic features of the MIT.
On increasing the hole density $d\rho /dT$ changes sign
%the temperature dependence of the resistivity 
%changes from $d\rho /dT<0$ to $d\rho /dT>0$ 
\cite{senz}. At the critical density $p_{c}$, we 
found $r_{s}=1/a_{B}^{\star}\sqrt{\pi p}\approx 8$, where $a_{B}^{\star}$ is the 
effective Bohr radius. The critical resistance was
$\rho_{c}\approx h/e^2$. The $r_{s}$ value is a measure of the 
importance of interaction effects and $\rho$ is a measure of the 
degree of disorder in the system.
Like in other systems temperature and electric
field scaling could 
be performed \cite{senz} allowing the interpretation in terms of a 
quantum phase transition. We have found in Ref. \cite{senz2}
that in the metallic regime
weak localisation (WL) reduces the metallic behaviour without destroying 
it, whereas localising interference corrections dominate the zero field resistivity 
in the insulating regime.

In the following we analyse magnetotransport data in the 
metallic regime by first extracting the contributions of
WL and interaction corrections to the temperature dependence 
of the conductivity; then the effect of temperature dependent 
screening is quantitatively discussed.

\begin{figure}
\noindent\epsfig{file=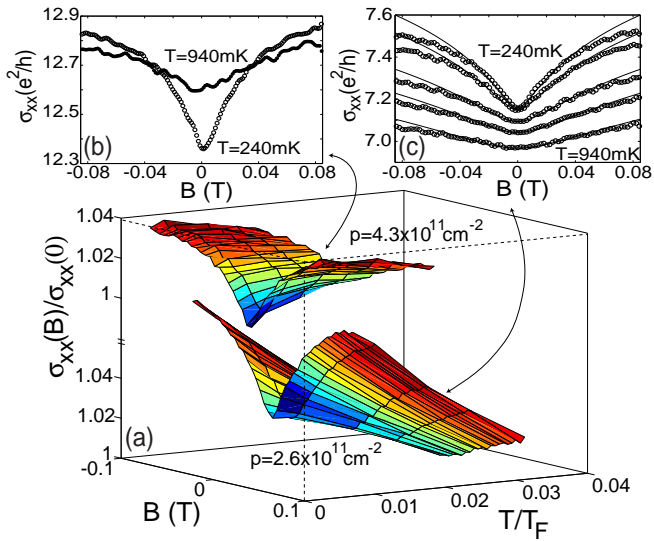,width=\linewidth,angle=0}
\centerline{\epsfxsize=3.5 in }
\caption{(a) Temperature and magnetic field dependence of the conductivity 
for densities $p=2.6\times 10^{11}$ cm$^{-2}$ (lower surface) and 
$p=4.3\times 10^{11}$ cm$^{-2}$
(upper surface). The upper surface shows reentrant insulating 
behaviour at lower temperatures. (b) Magnetoconductance for the 
higher density at the two extreme temperatures. (c) Magnetoconductance 
for the lower density at various temperatures.}
\label{fig5}
\end{figure}
In Fig. \ref{fig5}a we show magnetoconductivities measured 
at the highest densities $p=2.6\times 10^{11}$ cm$^{-2}$ (achieved 
under the gated part of the sample) and
$p=4.3\times 10^{11}$ cm$^{-2}$ (ungated part of the sample) for various
temperatures. Fig. \ref{fig5}b presents the magnetoconductivity curves
of the higher density for the lowest and highest temperature.
Fig. \ref{fig5}c shows magnetoconductivity curves of the lower density 
together with theoretical fits (solid lines) to be discussed later in 
this paper. It can be seen in Fig. \ref{fig5}c that 
the temperature dependence of the conductivity minimum around zero magnetic
field, which we attribute to the WL corrections, is 
overpowered by the much stronger temperature dependence of the 
background conductivity. We therefore analyse the 
zero-field conductivity $\sigma(T)$ in the spirit of Ref. \cite{inteff}
\begin{equation}
\sigma(T) = \sigma_{D}(T) + \delta\sigma_{WL}(T) + 
\delta\sigma_{I}(T),
\label{eq1}
\end{equation}
where $\sigma_{D}(T)$ is the Drude conductivity, $\delta\sigma_{WL}(T)$ 
and $\delta\sigma_{I}(T)$ are the WL and the interaction contributions, 
respectively. Strictly speaking, 
eq. (\ref{eq1}) is based on the validity of the Fermi-liquid 
description and $r_{s}\leq 1$ is required. Nevertheless we
empirically apply this concept to our system, since phenomenologically
the magnetoconductivity exhibits all the features also found in 
samples with low $r_{s}$.

In the well established theory of WL \cite{hikami}
spin-orbit relaxation 
mechanisms like the Elliott-Yafet mechanism, the Dyakanov-Perel 
mechanism or the Rashba-effect have been taken into account 
perturbatively \cite{hikami,soel}, which is appropriate for 
conduction band electrons. WL in
2D p-type systems such as p-GaAs has only recently been studied in 
detail \cite{pedersen}. In hole gases
the valence band is strongly influenced by spin-orbit 
interaction and strain. The spin relaxation time is of the same order 
as the momentum relaxation time and can therefore no longer be 
treated perturbatively \cite{golub}. It is therefore not {\em 
a priori} clear that the theories in Refs.
\cite{hikami,soel} can be applied to our system.

Here we apply the theory developed in Ref. 
\onlinecite{golub} for p-type 2D hole gases which takes the valence 
band structure
into account and includes the effect of HH-LH 
mixing.
Fig. \ref{fig5}c
shows theoretical curves fitted to the measured 
conductivity according to
\begin{eqnarray}
\delta&&\sigma_{WL}(B,T)-\delta\sigma_{WL}(0,T)=\frac{e^2}{2\pi^2\hbar}\times\nonumber \\
&&\left[f_{2}\left(\frac{B}{B_{\varphi}+B_{||}}\right)
+\frac{1}{2}f_{2}\left(\frac{B}{B_{\varphi}+B_{\perp}}\right)
-\frac{1}{2}f_{2}\left(\frac{B}{B_{\varphi}}\right)\right] 
\label{eqwl} 
\end{eqnarray}
Here $f_{2}(x)=\ln x+\psi(1/2+1/x)$, $\psi$ is the digamma function 
and
$B_{i}=\hbar / (4De\tau_{i})$ ($i = ||,\perp , \varphi$),
with the diffusion constant $D$ and the 
phase-coherence time $\tau_{\varphi}$. The other relaxation times 
are given in Ref. \onlinecite{golub} to 
be $1/\tau_{||}=1/\tau_{e}\cdot\left(k_{F}a/\pi\right)^4I_{||}$ and
$1/\tau_{\perp}=1/\tau_{e}\cdot\left(k_{F}a/\pi\right)^6I_{\perp}$, 
where $\tau_{e}$ is the transport relaxation time, $a$ is the width of 
the quantum well and $k_{F}$ is the Fermi wavevector.
The quantities $I_{||/\perp}$ depend on the ratio of the LH and 
HH mass and have to be computed numerically. In our samples
the holes are effectively confined by a triangular well due to the 
asymmetric doping and we estimate $k_{F}a/\pi\approx 0.2$.
We use $B_{\varphi},B_{\perp}$ and $B_{||}$ as fitting 
parameters and find that $B_{||},B_{\perp}\ll 
B_{\varphi}$, i.e. HH-LH mixing seems to be 
insignificant in our sample. This result is in contrast to the 
measurements on p-GaAs, where stronger HH-LH mixing 
leads to appreciable values for $B_{\perp}$ and $B_{||}$ \cite{pedersen}.
In our case eq. (\ref{eqwl}) reduces to the
result for negligible spin-orbit scattering \cite{hikami} already 
used in Ref. \cite{senz2}.
As it was evaluated there the phase breaking rate is
$1/\tau_{\varphi}\propto T$ and it 
increases with decreasing hole concentration as expected from theory. 

\begin{figure}
\noindent\epsfig{file=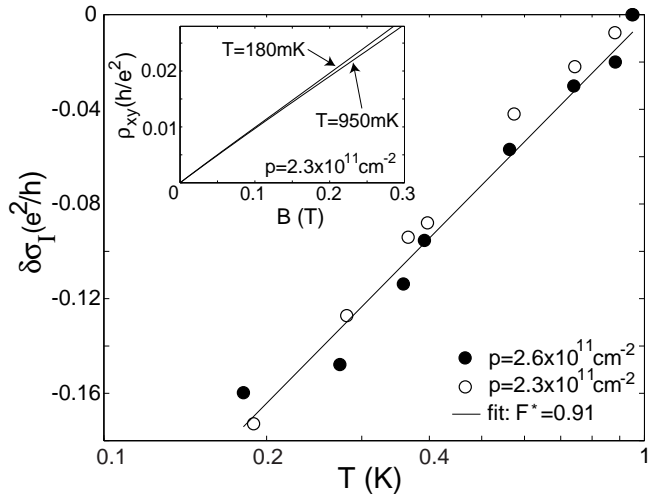,width=\linewidth,angle=0}
\centerline{\epsfxsize=3.5 in }
\caption{Interaction corrections $\delta\sigma_{I}(T)$
in the metallic regime. The solid line is a theoretical fit (see 
text). The inset shows the raw data for the highest and lowest 
temperature.}
\label{fig3}
\end{figure}

Interaction corrections to the conductivity can be extracted from the 
temperature dependence of the Hall-resistance at small magnetic 
fields \cite{poole}. It is predicted that the WL 
correction does not affect the Hall resistance, while interaction 
corrections obey \cite{inteff}
$ \delta\sigma_{I}=-\sigma\delta 
R_{H}/2R_{H}=e^2/(2\pi^2\hbar)(1-3F^\star/4)\ln(kT\tau_{e}/\hbar) $,
where $R_{H}=d\rho_{xy}/dB$ is the Hall constant. Figure 
\ref{fig3} shows $\delta\sigma_{I}(T)$ determined in this way for 
two different densities in the metallic regime.
We find that $\delta\sigma_{I}$ is negative and 
determine $F^\star=0.91$ in agreement with Ref. \cite{emeleus1}. This means that 
interaction corrections to the conductivity give another contribution 
of insulating behaviour to the total conductivity in the metallic phase. This 
result which agrees with Ref. \cite{cambridge} for p-GaAs is of special
importance, since interaction corrections have 
been suggested to lead to metallic behaviour.

\begin{figure}
\noindent\epsfig{file=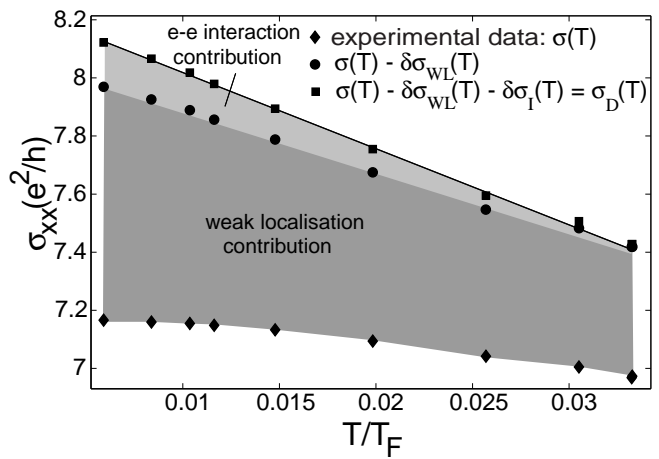,width=\linewidth,angle=0}
\centerline{\epsfxsize=3.5 in }
\caption{Zero-field contributions to the conductivity at $p=2.6\times 
10^{11}$ cm$^{-2}$. From the 
measured curve $\sigma(T)$ ($\blacklozenge$) we first subtract $\delta\sigma_{WL}(T)$
arriving at the curve marked $\bullet$ and then 
$\delta\sigma_{I}(T)$ which leaves us with the bare 
$\sigma_{D}(T)$ ($\blacksquare$).}
\label{fig4}
\end{figure}

Now that we have experimentally determined
$\delta\sigma_{WL}(T)$ and $\delta\sigma_{I}(T)$ we are 
in the position to extract the bare $\sigma_{D}(T)$ according to eq. 
(\ref{eq1}) by subtracting the two corrections from the measured 
zero-field conductivity (Fig. \ref{fig4}).
The obtained Drude conductivity shows a linear metallic 
temperature dependence which we will address in the following.

The temperature dependent Drude
conductivity is
\[
\sigma_{D}(T) = \frac{pe^2\tau_{e}(T)}{m^\star}
\]
where
$p$ is the sheet density of the holes and the average Drude 
scattering time $\tau_{e}(T)$ 
has to be calculated according to
\[ \tau_{e}(T) = \frac{\int dE\;E\tau_{e}(E,T)(-df/dE)}{\int 
dE\;E(-df/dE)}, \]
with the Fermi distribution function $f(E,T)$ and the energy dependent 
scattering rate
\[ \frac{\hbar}{\tau_{e}(E,T)} = 2\pi 
N_{i}\int\frac{d^2k'}{(2\pi)^2}\left|\frac{V(q)}{\varepsilon(q,T)}\right|^2(1-\cos\theta)\delta\left(E_{k}-E_{k'}\right) \]
Here $N_{i}$ is the density of ionized impurities, $V(q)$ is the 
matrix element for scattering by a wavevector 
$q=k_{F}\sqrt{2(1-cos\theta)}$ and $\varepsilon(q,T)$ is Lindhard's 
dielectric function.
For dominant large angle scattering (i.e. scattering for $q\approx 
2k_{F}$) it has been shown that \cite{dolgopolov}
\begin{equation}
    \sigma_{D}(T) = 
\sigma_{D}(0)\left[1-C(p)\frac{T}{T_{F}}\right]+{\cal O}\left[\left(\frac{T}{T_{F}}\right)^{3/2}\right].
\label{eqscr}
\end{equation}
The linear term stems from the temperature 
dependence of the dielectric function $\varepsilon(q,T)$, which 
becomes particularly important for large angle scattering. It is a 
direct consequence of electron-electron interactions causing a
temperature dependent polarizability of the 2D hole gas. Theory 
predicts values for the constant $C(p)$ which depend on the 
scattering mechanism and on the hole density $p$. If we apply eq. 
(\ref{eqscr}) to $\sigma_{D}(T)$ in Fig. 
\ref{fig4} we determine $C=3.1$ in reasonable agreement with a predicted 
value of $C\approx 2.8$ for charged interface impurity scattering at 
low $p$. Similar agreement has been found for all densities in the 
metallic regime. This demonstrates that temperature dependent screening can 
indeed explain the metallic temperature dependence of the resistivity 
in our p-SiGe system without invoking a novel metallic phase. It also 
implies that at sufficiently low temperatures insulating behaviour 
with $d\rho/dT<0$ is expected to be recovered. Similar results in 
p-SiGe were obtained for much higher electron densities in Ref. 
\cite{emeleus1}.

For higher carrier densities $p$ we expect $\sigma_{D}(T)$ to exhibit 
a weaker temperature dependence, since $C(p)$ decreases 
\cite{dolgopolov}. In Fig. 
\ref{fig5}a we therefore show the magnetic field and the temperature dependence 
measured at two densities. The lower surface shows the curves measured
at $p=2.6\times 10^{11}$ cm$^{-2}$. At large $\left| B\right|$ the metallic 
temperature dependence can be seen. At $B=0$ the WL 
peak tends to counteract this metallic behaviour without overpowering 
it in the range of temperatures shown. The upper surface was measured 
at a density of $4.3\times 10^{11}$ cm$^{-2}$ on an ungated part of the 
same sample. The temperature dependence at large $\left| B\right|$ is 
weaker than for the other surface ($C=0.8$), in agreement with temperature 
dependent screening. At $B=0$ however, the WL 
correction is strong enough to restore the insulating temperature 
dependence of the conductivity. Such a reentrant insulating behaviour 
is consistent with the observations made in Ref. \onlinecite{reentrant}.

In the following, we address the range of validity of our 
analysis.
The theories of linear screening, WL and interaction
corrections are expected to hold as long as the disorder in the 
system is low enough, i.e. as long as $k_{F}l>1$.
Therefore, our analysis is limited to the 
metallic regime. Non-linear screening models have to be invoked at
the MIT, where our system undergoes the transition from 
WL ($k_{F}l>1$) to strong localisation 
($k_{F}l<1$). We have empirically applied concepts developed for 
weakly interacting systems with $r_{s}\leq 1$. This attempt was 
successful and a Fermi-liquid description of the 2D hole gas appears 
to be consistent with the experiment, in spite of $r_{s}\approx 8$. 
The theoretical understanding and interpretation of this finding 
remains an open issue.

How relevant is this analysis for other systems in which the MIT was 
observed? Our analysis was simplified by the fact that 
large angle scattering is dominant in p-SiGe. For this special 
case, the analytical results of Ref. \onlinecite{dolgopolov} apply. 
Candidates for a similar analysis are therefore low-mobility 
Si-MOSFETS (for which this theory was originally developed) or the 
n-GaAs system with InAs quantum dots near the 2D electron 
gas \cite{ribeiro}. However, Kravchenko {\it et al.} stated in Ref. 
\onlinecite{kravchenko} explicitly that temperature dependent 
screening can not account for the metallic behaviour in their 
high-mobility Si-MOSFET samples. Performing a similar analysis on the 
n-type GaAs samples of Ref. \cite{ribeiro} we find that the 
temperature dependence in the metallic range is much too large 
compared to the theory used above. For p-type GaAs samples 
exhibiting the MIT temperature dependent screening was 
suggested in Refs. \onlinecite{cambridge,hamilton} as the relevant 
mechanism. 
Under the condition of small angle 
scattering this effect is much weaker than in our case.

In conclusion, we have analysed the magnetoresistance and the Hall 
resistance in p-SiGe samples in terms of interference and interaction 
corrections to the conductivity and found the measurements to be 
consistent with ordinary Fermi-liquid 
behaviour in spite of $r_{s}\approx 8$.
Both corrections tend to localise the system as the 
temperature is lowered. The temperature dependence of the background 
Drude-conductivity has been found to depend linearly on temperature. 
Its behaviour is in good agreement with the theory of temperature 
dependent screening for systems in which large angle scattering 
dominates. The analysis applies to densities where the system is
in the metallic regime and can not easily be extended to the 
transition region, where $k_{F}l\approx 1$.
Although these observations can not exclude the possibility of a 
novel metallic ground state in our or in other systems unambiguously, we are 
inclined to discard this exciting possible interpretation for p-SiGe 
hole gases on the basis of our experiments and analysis. The 
theoretical understanding of experimental consistency with
Fermi-liquid behaviour at these large $r_{s}$-values remains a 
challenging topic for further research.

We have enjoyed fruitful discussions with V. Dolgopolov and J.L. Pichard.
Financial support from ETH Zurich and the Schweizerischer 
Nationalfonds is gratefully acknowledged.

\end{multicols}
\end{document}